\documentclass[twocolumn,showpacs,preprintnumbers,amsmath,amssymb]{revtex4}
\usepackage{graphicx}
\usepackage{graphicx}
\usepackage{dcolumn}
\usepackage{bm}
\begin{document}
\title{An One Dimensional Adiabatic Model for Fusion Involving Loosely Bound and Halo Nuclei with
Heavy Targets}
\author{Ajit Kumar Mohanty}
\email{ajitkm@barc.gov.in}
\affiliation{Nuclear Physics Division, Bhabha Atomic Research Centre, 
Mumbai 400 085, India}
\date{\today}

\begin{abstract}
{
An one dimensional adiabatic model has been proposed for fusion of loosely bound and halo 
light nuclei with
various  heavy targets. It is shown that fusion cross sections at near and sub-barrier energies
can be explained using a simple WKB tunneling through an adiabatic barrier without invoking breakup
coupling explicitly. The model has been 
applied successfully to explain fusion cross sections for several systems including
recently measured $^{6}Li+^{198}Pt$ system (Phy. Rev. Lett. 103, 232702, 2009) where data exists well above and
below the barrier and $^{8}He+^{197}Au$ system (Phy. Rev. Lett. 103, 232701, 2009) where $^{8}He$ is 
highly neutron rich. Interestingly, the fusion of stable $^{4}He+^{197}Au$ system can not be
explained on the basis of this adiabatic model as it requires normal tunneling through the sudden
barrier. The requirement of adiabatic potential for the loosely bound and halo nuclei is linked to the
fact that for such systems neutron flow leading to neck formation is initiated at a larger distance 
which modifies
the sudden potential.  } 
\end{abstract}
\maketitle

Fusion cross section enhancement at sub-barrier energy over the prediction of a simple
barrier penetration model (BPM) is a wellknown phenomena for stable nuclei which occurs due
to the coupling of relative motion with the intrinsic degrees of freedoms of the projectile and target
nuclei \cite{bal,das1}. Similar enhancement has also been found for fusion of loosely bound and
halo light nuclei with various heavy targets \cite{kly}. However, unlike stable nuclei, the fusion 
cross sections at above Coulomb barrier energy shows about $10\%$ to $30\%$ suppression with respect
to the BPM predictions \cite{canto}.
 This could be  due to the presence of breakup components as the loosely bound projectiles
are more prone to breakup due to their low binding energy as compared to the stable counterparts. 
Two theoretical models having different perspectives have been proposed to understand
the role of breakup reaction on fusion process \cite{huss,tak,dasso}.
Intuitively, it can be told that increase of breakup process
may hinder fusion \cite{huss,tak}. On the otherhand, considering breakup process like any other reaction
channel, the coupled channel approach would lead to fusion enhancement at below barrier energy and a
suppression above it \cite{dasso,kly}. During last few years, high precision fusion cross section measurements
have been carried out for several systems including both loosely bound and halo projectiles like $^{6}Li$, $^{7}Li$,
$^{9}Be$ and $^{6}He$ with medium and heavy targets like $^{144}Sm$, $^{198}Pt$, $^{208}Pb$, $^{209}Bi$
and $^{238}U$ \cite{kolata,das2,agu,trotta,raabe,trip,das3,gomes,peni,rath,sri,lem}. Fusion enhancement over the
BPM predictions below the Coulomb barrier and suppression above it appears to be a generic phenomena \cite{canto}
with a
few exceptions \cite{trotta,raabe} where fusion cross sections might have contained a large component coming from
direct processes (incomplete fusion components). 
While improved CDCC type coupled channel calculations with inclusion of
breakup coupling have been used to explain 
the fusion enhancement at
below barrier energy, they all fail to explain the above barrier data unless $10\%$ to $30\%$ suppression factor
(almost energy independent)  is used \cite{kly}. In this letter, we propose a simple one dimensional model
where tunneling probability is estimated through an adiabatic barrier which has a shape thiner than the sudden potential
particularly in the nuclear interior region. It is shown that this adiabatic BPM can explain sub-barrier fusion enhancement
for several systems involving loosely bound light nuclei without invoking breakup coupling explicitly. Recently,
a new type of fusion hindrance has been observed at deep sub-barrier
energies for fusion involving stable nuclei \cite{jiang}. At sub-barrier energy, fusion cross section is enhanced
(over BPM calculations) as expected which can be explained by coupled channel calculations.
 However, when measurements are extended
to deep sub-barrier energies, fusion cross sections are suppressed with respect to the same 
coupled channel predictions which explains the data at sub and above barrier energies. Interestingly, the recent
fusion cross section measurements of $^{6}Li+^{198}Pt$ systems where data exists 
well above and below the barrier energies shows
no such deep sub-barrier hindrance \cite{sri}.
Similarly, there is another recent measurement of fusion cross section of $^{8}He+^{197}Au$ system which shows unusual
behavior of the tunneling of neutron rich $^{8}He$ nuclei as compared to normal $\alpha$ particle \cite{lem}.
We have shown here that the present adiabatic BPM  can also explain fusion cross sections of the above two systems
without invoking any channel coupling mechanism explicitly.

\begin{figure}
\begin{center}
\includegraphics[scale=.4]{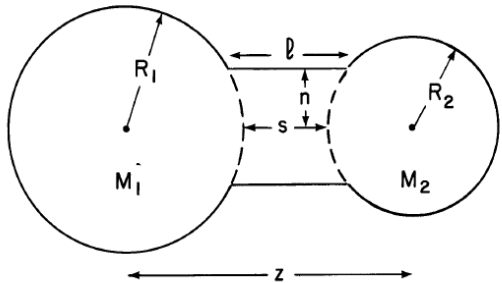}
\caption{A macroscopic representation of two nuclei of mass $M_1$ and $M_2$ and connected by a cylindrical neck
of length $l$, radius $n$ and surface to surface distance $s$. 
Note that the centre to centre distance $z$ and the radial distance $r$ which is used in the text  have the similar
meaning.}
\label{fig1}
\end{center}
\end{figure}

\begin{figure}
\begin{center}
\includegraphics[scale=.4]{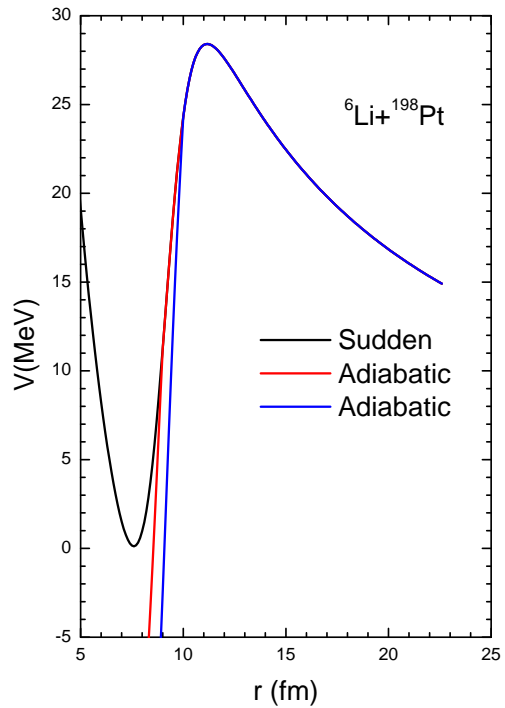}
\caption{The potential $V$ as a function of inter nuclear distance $r$ for a typical
$^{6}Li+^{198}Pt$ system plotted using the parameters as listed in table I. The black
 curve represents the
sudden potential where as the red and blue curves are adiabatic potentials for adiabatic parameter
$R_a$ =$9.0$ fm and $10.0$ fm respectively. For detail see the text.}
\label{fig2}
\end{center}
\end{figure}

\begin{figure}
\begin{center}
\includegraphics[scale=.4]{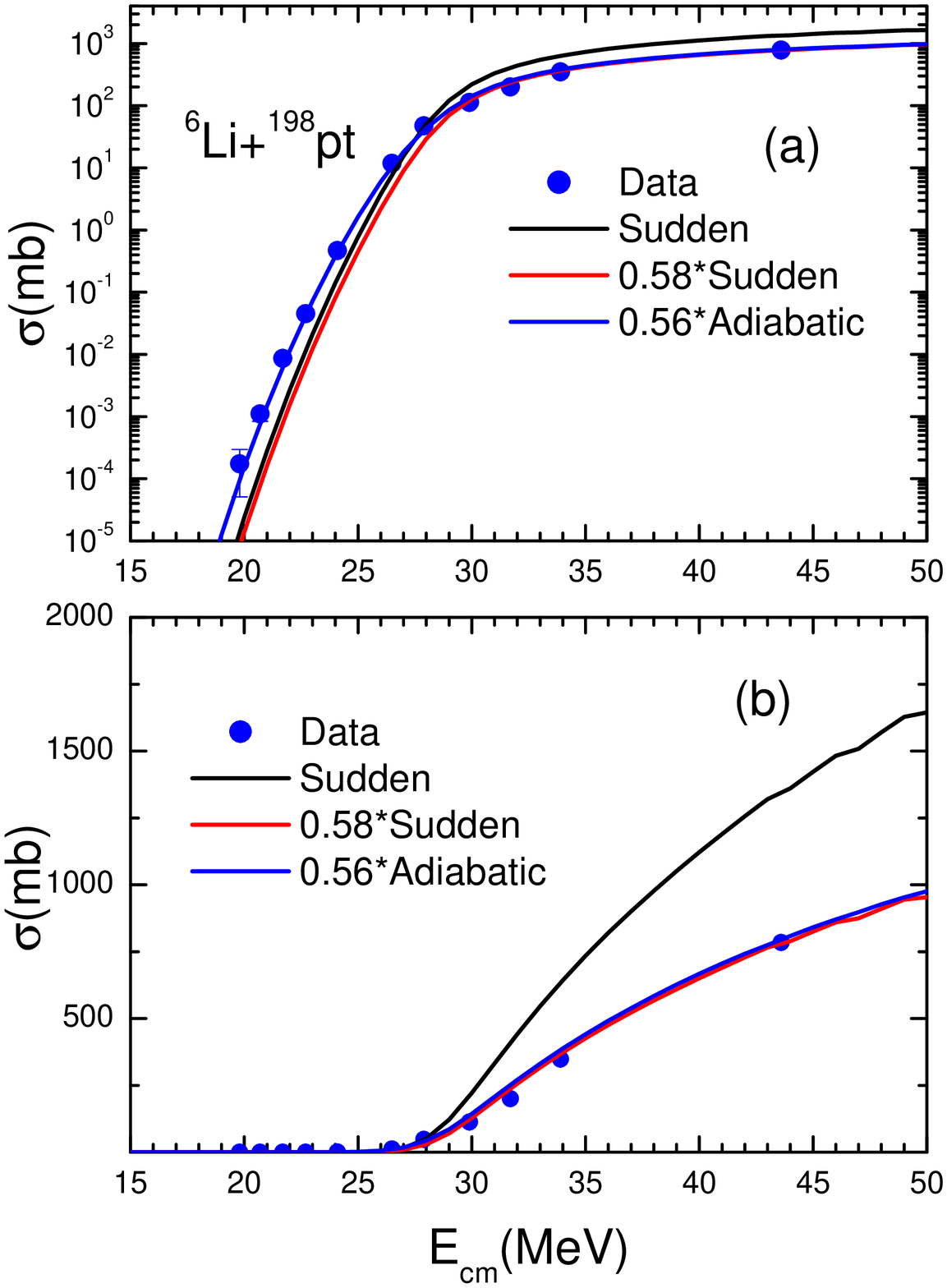}
\caption{(a) Fusion cross section for $^{6}Li+^{198}Pt$ system as a function of energy. The black curve
is obtained using only the sudden potential while the red curve is obtained with a suppression
factor 0f $0.58$. The blue curve is obtained using adiabatic
 potential with a suppression factor of $0.56$. (b) The plots are shown in the linear scale.
The data points are taken from \cite{sri}.}
\label{fig3}
\end{center}
\end{figure}

We construct an adiabatic potential from the commonly used sudden potential by adding an extra correction term which
arises due to the neck formation and is given by,
\begin{eqnarray}
V(r,n)=V_N(r)+\frac{Z_1Z_2e^2}{r}+V_{neck}(r,n)
\label{vad}
\end{eqnarray}
where $V_N$ is the sudden ion-ion nuclear potential, $r$ is the centre to centre distance between the two nuclei
and $n$ is the neck parameter (defined below) which is meaningfull only at the nuclear overlap region being characterized
by an adiabatic  distance $r=R_a$. The strong nucleon exchange (nucleon flow) between two
nuclei (involving loosely bound and halo nuclei) at $r>R_1+R_2$ is well manifested, for example,
 by the sub-barrier fusion enhancement of $^{6}He+^{208}Pb$ system with sequential neutron transfer from $^{6}He$
to the $Pb$ nucleus with positive $Q$ value \cite{peni,zag}. An important aspect of the nucleon exchange is that it
provides the formation of an intermediate di-nuclear state in the fusion reaction before the two nuclei actually
fuse. We model this aspect  using a macroscopic approach where two nuclei are  connected by a cylindrical neck as
shown in Fig. \ref{fig1}. The extra energy due to neck formation is proportional to the
surface area of the cylinder of length $l$ and radius $n$ and can be written as \cite{swiat},
\begin{eqnarray}
V_{neck}(r,n) \approx 2 \pi \gamma (n s -n^2+\frac{n^3}{2 \bar R})
\label{n1}
\end{eqnarray} 
where $s=r-(R_a-\bar R/2)$, $\bar R=(R_1 R_2)/(R_1+R_2)$ and $\gamma$ is the surface tension co-efficient
$\sim 1.0$ MeVfm$^{-1}$. The distance $R_a$ is called the adiabatic parameter that decides at which point neck opening
becomes favorable  \cite{note}. Using the dimensionless variables $\rho=s/(2\bar R)$ and $\nu=n/(2 \bar R)$, 
Eq. \ref{n1}
can be written as,  
\begin{eqnarray}
V_{neck}(\rho,\nu) = 8 \pi \gamma \bar R^2(\rho \nu -\nu^2+\nu^3)
\label{neck}
\end{eqnarray} 
The above expression is a simple cubic order polynomial that vanishes at $\nu=0$ and has a minimum at
$\bar \nu=(1+\sqrt{1-3\rho})/3$ as long as $\rho \le 1/3$. For neck relaxation, it is required for the
di-nuclear system to move from $\nu=0$ to a state characterized by $\nu =\bar \nu.$ 
It can be seen that for $1/4 \le \rho \le 1/3$, the minimum
at $\nu=\bar \nu$  is always higher than the value at $\nu=0$. Since this minima corresponds to a metastable sate,
neck relaxation is not favored. 
For $\rho=1/4$, $V_{neck}(\nu=0)$ and $V_{neck}(\nu= \bar \nu)$ are degenerate (equal to zero) and
 $V_{neck}(\bar \nu)$ becomes less than $V_{neck}(\nu=0)$ (becomes negative) 
for $\rho \le 1/4$. Hence, neck relaxation is possible for $\rho \le 1/4$  corresponding to $r \le R_a$.
Therefore, we set $V_{neck}(\bar \nu)=0$ for $r>R_a$ and evaluate it at $\nu=\bar \nu$ using Eq. \ref{neck}
for $r\le R_a$. Finally, we estimate the adiabatic potential from Eq. \ref{vad} by adding the above neck
potential at $\nu=\bar \nu$. As expected, the potential given in Eq. \ref{vad} is of sudden nature for $r>R_a$ and becomes
adiabatic for $r\le R_a$. Although, we treat $R_a$ as parameter, 
we expect it to lie in between  $R_1+R_2$ and $ R_b$ where $R_b$ is the radius of the
Coulomb barrier. For the nuclear part $V_N$, we use the Akuyz-Winther(AW) parameterization   
given by \cite{esben}
\begin{eqnarray}
V_N(r)=\frac{-16 \gamma~\bar R~a}{1+exp\left \{\left(r-R_1-R_2-\Delta R\right)/a
         \right \}},
\end{eqnarray}
where $\Delta R$ is an adjustable parameter used to reproduce the Coulomb barrier. 
Here, $\gamma=0.95$ MeV/fm$^2$ is the nuclear surface tension co-efficient , 
$R_i=1.2A_i^{1/3}-0.09$ fm, the diffuseness
parameter $a=0.63$ fm,  and $\bar R = R_1R_2/(R_1+R_2)$. 

Fig. \ref{fig1} shows the plot of total nuclear potential as a function of $r$ for 
$^{6}Li+^{198}Pt$ system as
an example. The black curve is the sudden potential without any 
neck correction. The red and blue curves
are the adiabatic potentials for two different values of $R_a$. 
As discussed before, the adiabaticity begins
for $r\le R_a$ and the adiabatic potential becomes thiner as compared to its sudden counter part. Using
the above potential, we now estimate the fusion cross section using 

\begin{equation}
\sigma_f=\sum_l \sigma_l=\frac{\pi}{k^2}\sum_l (2l+1) T_l(E),
\label{fus}
\end{equation}
where $k$ is the relative wave number and $T_l(E)$
is the tunneling probability which can be estimated using the WKB approximation,
\begin{eqnarray}
T_l(E)=\frac{1}{ 1+ exp(2S_l) },
\label{wkb}
\end{eqnarray}
where $S_l$ is the classical action given by,
\begin{eqnarray}
S_l = \frac{2\mu}{\hbar^2} \int_{r_1}^{r_2} \sqrt{\left (V_l(r)-E \right)} dr,
\end{eqnarray}
and 
\begin{eqnarray}
V_l(r)= V_C(r)+V_N(r)+\frac{l(l+1) \hbar^2}{2\mu r^2}.
\end{eqnarray} 
Under the parabolic approximation, Eq.(\ref{wkb}) can also be estimated using Hill-Wheeler
expression \cite{hill},
\begin{eqnarray}
T_l(E)=\left [ 1+ exp\left (\frac{2\pi}{\hbar \omega} (V_b^l-E) \right ) \right ]^{-1}.
\label{hill}
\end{eqnarray}
where 
\begin{eqnarray}
V_b^l= V_b+\frac{l(l+1) \hbar^2}{2\mu R_b^2}.
\label{bar1}
\end{eqnarray} 
and $V_b$ being  the $s$-wave barrier.
Finally, we estimate tunneling probability using WKB approximation (Eq. \ref{wkb}) 
for sub-barrier energy and Eq. \ref{hill} for energy above the Coulomb
barrier. Although $R_a$ is a variable, it is noticed that best result is obtained when
$R_a$ is close to $R_b$ (with a few exception as listed in table I).
The second
parameter $\Delta R$ of our model is fixed to reproduce the Coulomb
barrier for various systems which are taken from the literatures. In many cases, the Coulomb
barrier have been determined in a model independent way by estimating the 
centroid of the experimental barrier distributions \cite{das2,das3,gomes}.    

\begin{figure}
\begin{center}
\includegraphics[scale=.4]{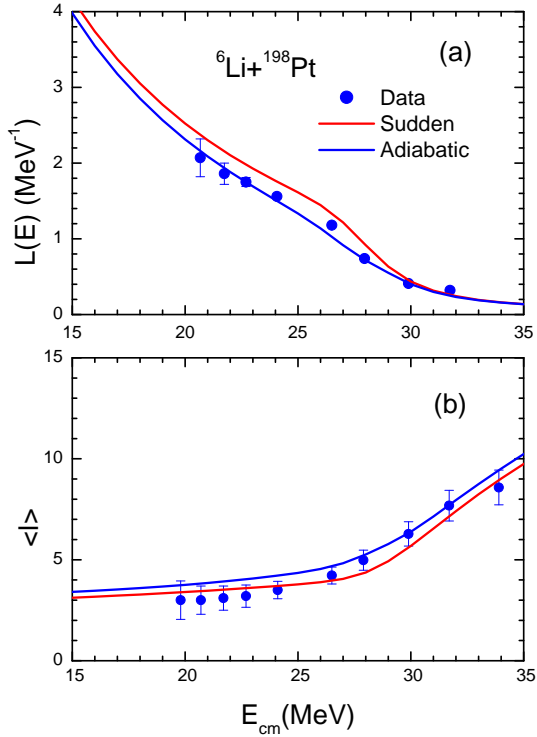}
\caption{The $(L(E)$ and $<l>$ as a function of energy $E_{cm}$ for $^{6}Li+^{198}Pt$ system both
for sudden (red curve) and adiabatic (blue curve) potentials. Data points are taken from \cite{sri}.}
\label{extra}
\end{center}
\end{figure}

\begin{figure}
\begin{center}
\includegraphics[scale=.4]{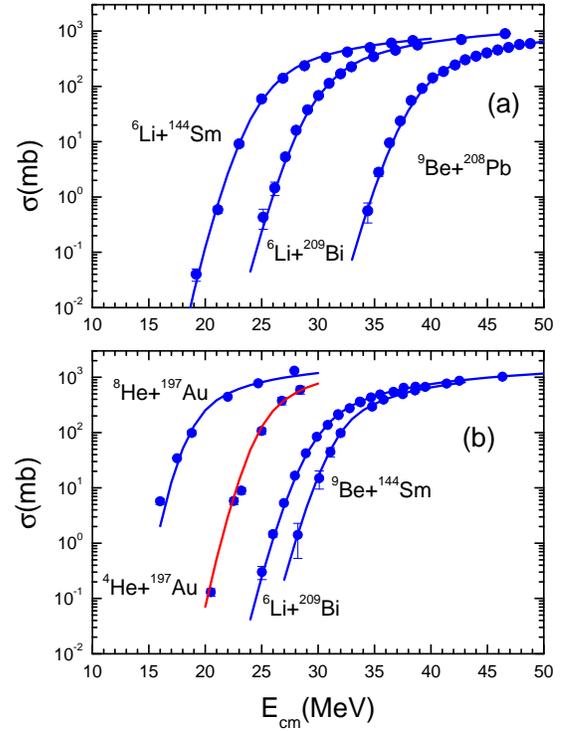}
\caption{Fusion cross section versus energy for different projectile and target
combinations as shown in figure captions using adiabatic model except for
$^{4}He+^{197}Au$ system which is obtained using only sudden potential (red curve). The X-axis
for $^{4}He+^{197}Au$ system has been shifted by $+5$ MeV for clarity. Experimental 
data points are taken from \cite{gomes,rath,lem,basu}.
}
\label{fig4}
\end{center}
\end{figure}

We estimate fusion cross section using barrier penetration model both for sudden and adiabatic potential. 
The black curve in Fig. \ref{fig3} shows the BPM calculations using sudden potential  for $^{6}Li+^{198}Pt$
system. As expected, the experimental data are suppressed at above barrier energies and enhanced at sub-barrier
energies as compare to the predictions of the BPM calculations.  The linear scale in Fig.\ref{fig3}b shows
the above barrier suppression more prominently. The red  curve is  obtained when the
BPM result is scaled down by a factor of $0.58$ to explain the above barrier suppression where as the
results are not affected much at sub-barrier energies. Next, we carry out BPM calculations using adiabatic
potential  with  $R_a \sim 11.0$ fm  which is quite close to $R_b \sim 11.2$ fm . The resulting
blue curve  with a similar suppression factor  
explains the experimental data  quite well both at above and below
the barrier energies. It may be mentioned here that the  
coupled channel calculations (with inclusion of breakup channel) which has been used in \cite{sri} 
to explain $^{6}Li+^{198}Pt$ data also requires a similar suppression factor to explain the above
barrier data. Whether it is BPM or coupled channel calculations, the above barrier suppression seems to be
a generic feature and the fusion cross section needs to be scaled down by almost a constant factor whose
magnitude may depend on the specific model used. Fig.\ref{extra} shows the plot of logarithmic derivative $L(E)=
d[log(\sigma E)]/dE$ and average angular momentum $<l>$ as a function of energy for the same $^{6}Li+^{198}Pt$
system. The BPM results with adiabatic potential explains the data quite well (blue curves). It is
interesting to note that the BPM calculations with sudden potential also explains $<l>$ measurements although  
it fails to explain sub-barrier fusion enhancement and $L(E)$ behavior (see red curves).

We have applied this adiabatic model to few other systems. 
Fig.\ref{fig4}a shows the fusion cross sections for
$^{6}Li+^{144}Sm$, $^{6}Li+^{209}Bi$, $^{9}Be+^{208}Pb$ systems where as  Fig.\ref{fig4}b shows
the results for  $^{8}He+^{197}Au$, $^{7}Li+^{209}Bi$ and $^{9}Be+^{208}Pb$ systems obtained using
barrier penetration model with adiabatic potential except for $^{4}He+^{197}Au$ system which does not need any
adiabatic correction. The experimental data for the last system
can be explained using a BPM model with sudden potential and also without
any suppression factor. The potential parameters used in the calculations are listed in table I. As mentioned
before, we first adjust the $\Delta R$ parameter to reproduce the Coulomb barrier $V_b$ (second column) which are
taken from the literatures. The resulting $R_b$ and $\hbar \omega$ values are listed in column 4 and 5. The
bracketed values in column 4 shows the $R_a$ parameter which has been used to estimate the adiabatic potential.
The last column shows the suppression factor $f$ which is required to explain the above barrier data. Again
the values in the bracket shows the suppression factors which are required to explain the above barrier data
using BPM with only sudden potential. Since the adiabatic potential is thiner than the sudden
potential, the adiabatic model slightly overpredicts the fusion cross sections even at above barrier
energies as compared to predictions of the BPM 
with sudden potential. Therefore, the $f$ factors systematically turns out to be slightly lower than BPM
predictions with sudden potential. Note that, fusion with $^{4}He$ nuclei does not require any suppression factor
($f=1$) nor requires any adiabatic correction. 

In conclusion, it is shown that a simple barrier penetration formalism with adiabatic potential (a potential which
is thiner than the sudden potential in the nuclear interior region) can explain sub-barrier fusion enhancement
of loosely bound and halo nuclei with heavy targets without invoking any breakup coupling explictly.
The excellent agreement between the BPM calculations and the experimental measurements suggests that the breakup
effect which is more important at above barrier energy, can be simulated by a suppression factor
 which is practically energy independent. Apart from breakup process, presence of other rotational and vibrational
states may affect the fusion process which has not been considered in the present BPM formalism. It will be ideal to
use normal couple channel formalism (without any breakup effect) with a bare potential which is adiabatic in nature.
Although such a model will explain sub-barrier fusion enhancement, 
the above barrier measurements will still require a 
suppression factor $f$ which may turn out to be slightly higher than what is shown in table I as coupled
channel calculation has some amount of inherent
suppression built in. It may be mentioned here that the large value of $R_a$ indicates that neck formation
becomes effective around the Coulomb barrier ($R_a \sim R_b$) due to neutron flow which may be a meaningfull
proposition for halo and loosely bound nuclei as contrast to the stable system like
$^{4}He+^{197}Au$ which does
not require any adiabatic correction.  Although the model used here is based on a simple BPM picture which works
well for light nuclei, the message which we want to convey is that the basic ion-ion
potential may become adiabatic in nature for halo and loosely bound nuclei. This aspect should not be
neglected while carrying out a proper couple channel
calculation.

\begin{table}
\caption{The potential parameters for various systems used in the calculations. The only
parameter $\Delta R$ has been adjusted to reproduce the Coulomb barrier which are taken from the
literatures.} 
\begin{tabular}{|l|c|c|c|c|c|}\hline\hline
System  &~~~~$\Delta R$ & ~~~~$V_b$ &~~~~ $R_b(R_a)$ &~~~~$\hbar \omega$~~~& f \\ 
\hline
$^{6}Li+^{198}Pt$  & 0.11 & 28.4 & 11.2 (11.0)& 5.0 & 0.56(0.58)  \\
\hline
$^{6}Li+^{144}Sm$  & -.41 & 25.2 & 9.9(9.9) & 5.0 & 0.60(0.64)  \\
\hline
$^{6}Li+^{209}Bi$  & 0.07 & 30.1 & 11.2(11.2) & 5.1 & 0.60(.64)  \\
\hline
$^{7}Li+^{209}Bi$  & 0.06 & 29.7 & 11.4(11.3) &4.7 & 0.68(0.72)  \\
\hline
$^{9}Be+^{208}Pb$  & 0.18 & 38.5 & 11.6(11.5) &4.7 & 0.64(0.67)  \\
\hline
$^{9}Be+^{144}Sm$  & 0.12 & 31.2 & 10.7(10.4) &4.4 & 0.80(0.85)  \\
\hline
$^{8}He+^{197}Au$  & -.10 & 18.7 & 11.5(11.5) &3.5 & 0.71(0.80)  \\
\hline
$^{4}He+^{197}Au$  & 0.10 & 19.8 & 10.8 &5.2 & 1.0  \\
\hline
\end{tabular}
\end{table}

\end{document}